\documentstyle[preprint,aps,amssymb]{revtex}  

\begin{document}
\draft                       


\title{The glass transition in a model silica glass: \\
       evolution of the local structure}

\author{P.~Jund and R.~Jullien}

\address{Laboratoire des Verres - Universit\'e Montpellier 2\\
         Place E. Bataillon Case 069, 34095 Montpellier France
	}


\maketitle


\begin{abstract}
We use  molecular dynamics simulations and the Vorono\"\i\
tessellation to study the geometrical modifications as a function of
temperature in a model silica glass. 
The standard deviation of the cell volumes, which is a measure of
the local density fluctuations, decreases with decreasing temperature,
as if it would like to vanish  at zero temperature. This evolution 
towards an ordered state is frozen out at the glass transition and 
consequently an amorphous sample is obtained at low temperature.
This structural freezing following upon the glass transition is noticeable 
in all the other geometric characteristics of the Vorono\"\i\ cells 
and a possible interpretation in terms of geometrical frustration is
proposed.
\end{abstract}



\narrowtext

\section{ Introduction }

Silica is a common material which is of great importance in chemistry,
geology and industrial applications. It is also a prototype of a network
forming glass. All these reasons explain why it has been the topic
of a great amount of studies. Nevertheless many features of this
typical ``strong'' glass need still to find a satisfactory explanation.
For example the origin of the First Sharp Diffraction Peak (FSDP) is still
controversial (Elliott 1991; Gaskell and Wallis 1996; Fayos, Bermejo, Dawidowski, Fischer and Gonzalez 1996). The origin (and the connection to the FSDP) of
the so-called Boson peak remains the topic of many studies, both 
experimental and theoretical (Shuker and Gammon 1970, Martin and Brenig 1974;
Akkermans and Maynard 1985; Buchenau, Zhou, N\"{u}cker, Gilroy and Phillips
1988; Malinovsky, Novikov, Parshin, Sokolov and Zemlyanov 1990;  Novikov and
Sokolov 1991; Sokolov, Kisliuk, Soltwisch and Quitmann 1992; B\"orjesson, Hassan, Swenson, Torell and Fontana 1993; Gurevich, Parshin, Pelous and Schober 1993; Achibat, Boukenter and Duval 1999; 
Schirmacher and Wagener 1993; Bermejo, Criado and Martinez 1994; Terki, Levelut, Boissier and Pelous 1996). With the development of the computing
speed a new type of studies has emerged in the past decade which can be
called ``computer experiments''. Indeed numerical simulations have been
used to study the vibrational spectrum (Jin, Vashista, Kalia and Rino 1993; Taraskin and Elliot 1997a; Horbach, Kob and Binder 1997; Taraskin and Elliott 1997b; Guillot and Guissani 1997) or 
the structural characteristics (Woodcock, Angell and Cheeseman 1976; Holender and Morgan 1991; Della Valle and Andersen 1992; Servalli and Colombo 1993; Nakano, Bi, Kalia and Vashista 1993; Sarnthein, Pasquarello and Car 1995; Vollmayr, Kob and Binder 1996)
of various model silica 
glasses. Within this framework we present in this paper a classical molecular 
dynamics study essentially focused on 
the evolution of the local structure in such a system. \\
The first step in that kind of endeavor is to choose the 
interparticle potential: we decided to use the two-body potential 
proposed by van Beest et al. (van Beest, Kramer and van Santen 1990). 
Indeed a study of the influence 
of the quenching rate on the properties of amorphous silica has 
shown that this potential based on {\em ab initio} parameters gives 
excellent results compared to experimental data both for
the structural characteristics (Vollmayr et al. 1996) and the 
vibrational properties (Taraskin and Elliott 1997b)
of vitreous silica, even if, in simulations the quenching rates are always
faster than in real systems . Similarly to what has been done earlier 
in the case of high pressure silica samples (Rustad, Yuen and  Spera 1991) 
we combine the molecular 
dynamics with the Vorono\"\i\ tessellation scheme in order to have a 
better insight into the structural evolution of the system during the 
quenching procedure. We want to know if and how the glass transition
leaves a ``signature'' in the geometrical characteristics of the Vorono\"\i\
cells. We recall that the Vorono\"\i\ cell attached to a particle 
is an extension, for disordered systems, of the Wigner-Seitz cell and gives
information on the local structure around this particle.\\
Our results as a function of temperature show the structural freezing
consecutive to the glass transition in all the geometric characteristics
of the Vorono\"\i\ cells.  
We suggest that the structural freezing below $T_g$ is mainly 
a consequence of the impossibility to build an ideal structure as a 
consequence of the so-called ``geometrical frustration''
(Kleman and Sadoc 1979; Kleman 1983, 1985, 1989; Sadoc and Mosseri 1997).\\
In section II a description of the simulations will be given. 
We present  our results in section III and discuss them in section IV.
Finally in section V we present our conclusions.


\section{Simulations}

If one wants to perform realistic simulations on silica glasses 
the major point is to choose an interaction potential which gives 
reasonable results compared to experimental data. Several choices 
are today possible, but in the last years one of the most successful 
classical potentials is the so-called BKS potential developed by 
van Beest et al. (van Beest et al. 1990). Though designed originally from  the 
crystalline phases of silica, it has been shown that it also 
describes very well the structural (Vollmayr et al. 1996) and vibrational 
(Taraskin and Elliott 1997b) properties of amorphous silica.\\
The functional form of the pairwise BKS interaction between two particles 
$i$ and $j$ is given by
\begin{equation}
U(r_{ij}) = \frac{q_iq_je^2}{r_{ij}} + A_{ij}\exp(-B_{ij}r_{ij})
            - \frac{C_{ij}}{r_{ij}^6}
\end{equation}
where $r_{ij}$ is the interparticle distance, $e$ the charge of an electron
and the parameters $A_{ij}$, $B_{ij}$ and $C_{ij}$ are fixed as follows:
$A_{SiO}=18003.7572$ and $A_{OO}=1388.773$ eV;$B_{SiO}=4.87318$ and $B_{OO}=2.76$ \AA$^{-1}$; $C_{SiO}=133.5381$ and $C_{OO}=175.0$ eV\AA$^6$. Note that
except for the Coulomb interaction ($q_{Si} = 2.4$ and $q_{O} = -1.2$) there is
no interaction between Si atoms.\\
This original form contains an unphysical property at short distances since
it diverges to minus infinity. To overcome this drawback which is especially
annoying at high temperature, we have added a short range repulsive term
($\sim 1/r_{ij}^{40}$) which insures that the potential goes to 
infinity at small interatomic distances and is practically 
equivalent to the original potential for $r_{ij} \geq 1.2$\AA\ 
(Si-O interaction), and $r_{ij} \geq 1.6$\AA\ (O-O interaction) without
introducing any artificial energy barrier.\\
The Coulomb interactions were computed using the Ewald summation
method (Ewald 1921; Allen and Tildesley 1990) with a characteristic 
constant $\kappa = 6.5/L$, 
where $L$ is the cubic box size, and considering 501 k-vectors in reciprocal 
space ($|k| \leq 6\times2\pi/L$). These values insure that the potential 
energy is obtained with a relative error smaller than $5\times10^{-4}\%$. 
No cut-off was used for the pairwise interaction.\\
We performed molecular dynamics simulations for microcanonical systems
containing 216 silicon and 432 oxygen atoms confined in a cubic box
of edge length $L = 21.48 \AA$, which corresponds to a mass density of
$\approx 2.18$g/cm$^3$ very close to the experimental value of $2.2$g/cm$^3$. 
Periodic boundary conditions were used to limit 
surface effects. In order to insure energy conservation even at high
temperature a timestep of $0.7$ fs was necessary. This value is substantially
lower than the one used in previous studies ($1.6$ fs (Vollmayr et al. 1996) or
$1.0$ fs (Taraskin and Elliott 1997b)) which may be due to our 
``conservative'' potential
correction. The 4-th order Runge-Kutta algorithm was used to integrate
the equations of motion.\\
The glass configurations were obtained by quenching well equilibrated
initial liquid samples obtained by melting $\beta$-cristobalite 
crystals at a temperature around 7000 K. After full equilibration of
the liquid ($\approx 40000$ timesteps), the system was cooled to 
zero temperature at a quench rate of $2.3\times10^{14}$ K/s which was obtained
by removing the corresponding amount of energy from the total energy of 
the system at each iteration. Due to computer time limitations 
this cooling rate is rather fast but it
has the advantage compared to other procedures involving either stepwise
cooling (Vollmayr et al. 1996) or temperature dependent rates (Taraskin and Elliott 1997b), to be linear and continuous all along the quenching procedure. At 
several temperatures during the quenching process the configurations 
(positions and velocities) were saved.
Each configuration was used to start a constant-energy molecular dynamics
calculation during which the temperature was recorded as a function of time.
The temperature was in all cases remarkably stable and only slight relaxation
effects could be observed. Nevertheless to avoid transient configurations
we allowed for each temperature and for each sample, 10000 relaxation 
steps followed by 50000 supplemental time steps (for a total simulation time
of $42$ ps) during which all the calculations were done. Apart from the
calculation of standard quantities (radial pair distribution 
function, mean square displacement) we included also in our molecular dynamics code a Vorono\"\i\ tessellation scheme similar to the one that we have 
developed for monocomponent soft-sphere glasses (Jund, Caprion and Jullien 1997a; 1997b). 
This scheme has been modified to take into account several types 
of atoms and thus it permits to follow the local structure around 
the silicon atoms and the oxygen atoms as a function of temperature during 
the quenching procedure. 
Here the Vorono\"\i\ cell is always defined as being the region of space 
closer to a given atom center than to any other and no dissymetry between the 
two components has been introduced as it is the case in the "navigation map" 
procedure (Gellatly and Finney 1982).
The Vorono\"\i\ cell characteristics as well
as all the other quantities have been averaged over samples 
obtained from 5 independent starting configurations in order to improve the
statistics of the results. The whole simulation lasted for more
than 3 million timesteps which were run on 4 nodes of 
an IBM/SP2 parallel computer.

\section{Results}

As said earlier this potential has already been used in other 
studies of amorphous silica and since we did not modify the BKS parameters
our results concerning the radial pair distribution (Vollmayr et al. 1996; 
Taraskin and Elliott 1997b) or the
diffusion constant (Guillot and Guissani 1997) are exactly identical 
to the referenced results and therefore we do not come back to 
these standard results here. Our aim is
to localize the glass transition temperature $T_g$ through the study of the 
structural characteristics (via the Vorono\"\i\ tessellation) of our model 
silica system. Nevertheless a straightforward way of determining $T_g$
is to monitor the potential energy versus the temperature, as has 
already been done for another model silica glass (Della Valle and 
Andersen 1992). The evolution
of the average potential energy per particle versus the temperature
is shown in Fig. 1. With increasing temperature, the potential energy
as well as the standard deviation increases as expected. Nevertheless
one can observe an acceleration of this increase between $3000$ and $4000$ K,
corresponding to the passage from a solid to a liquid behavior. 
Due to the fast cooling rate the value of $T_g$ is much higher than 
the experimental value ($1446$ K (Angell 1988)) but it is coherent 
with the value of $\approx 3500$ K obtained from the fit of $T_g$ versus the 
quenching rate proposed by Vollmayr et al. (Vollmayr et al. 1996). 
Note that we observe
a non negligible increase of the potential energy in the glass phase 
contrarily to what was obtained by Della Valle et al (Della Valle and 
Andersen 1992, Fig. 3)
who used a direct minimization procedure after the quench to investigate
the ``inherent structures'' (Stillinger and Weber 1982). Here we let the 
system evolve freely after the quench and it seems that even at low 
temperatures structural relaxation occurs. Note  also  that for such a
fast cooling rate the structure of the resulting silica might very well 
be slightly different than the one observed experimentally, since it is 
known that to get the most ``ideal'' glass, one should use the slowest 
quenching rate which does not produce crystallization (Kauzman 1948; 
Jund, Caprion and Jullien 1997c). However we recall that previous 
authors (Vollmayr et al. 1996;  Taraskin and Elliott 1997b), 
using the same potential with a quenching rate of the same order of 
magnitude as ours, were able to reproduce several physical properties 
of real silica.\\
Once we know approximately the value of $T_g$ (to our purpose this level
of accuracy is sufficient) we can tackle the study of the geometrical 
characteristics of the Vorono\"\i\ cells in order to follow the local
structure as a function of temperature. All the characteristics of the cells 
have been obtained (surface, number of faces, number of edges, etc...) but
we want to discuss here only some representative quantities. The first
one is the variation of the volume of the Vorono\"\i\ cells. This variation
is represented in Fig.2 for the silicon (a) and the oxygen (b) atoms together
with the corresponding standard deviations (c). With decreasing temperature
the volume of the silicon cell decreases while the volume of the oxygen cell
increases (these opposite variations are a consequence of our constant-volume 
calculations). Again a change of behavior is visible  and
corresponds to a slowing down of the evolution below the glass 
transition temperature.
An even more striking behavior is observed
in Fig.2c where the standard deviation $\sigma_V$ is plotted as a 
function of temperature. This is a quantity of physical interest 
since it measures the local density fluctuations around the particles. 
For both types of atoms, $\sigma_V$ decreases with decreasing temperature
as if it would like to tend to zero at $T=0$K but then below
$4000$ K this trend is stopped and finally $\sigma_V$ saturates around non-zero
values characteristic of spatial disorder. This is a direct observation of 
the low-temperature saturation of the density fluctuations which is a 
signature of the glass transition. Even though not reported here, 
exactly the same behavior can be
observed for the evolution of the surface of the Vorono\"\i\ cells as a
function of temperature.\\
To investigate further the structural evolution during the quench, 
we looked at the angle distributions. To numerically determine the bond angle 
O-Si-O (resp. Si-O-Si), we determine for each Si (resp. O) atom the two 
nearest O (resp. Si) atoms and we calculate the angle between the two 
corresponding segments Si-O (resp. O-Si), the result being averaged over 
all the Si (rep. O) atoms.
First we studied the tetrahedral 
O-Si-O angle which should be ideally equal to $109^\circ.47$ in a perfect 
tetrahedron. As can be seen in Fig.3a, this angle varies between  
$110^\circ.5$ at low temperature and $117^\circ$ at $7000$ K, 
with a slight change of behavior around $T_g$.
The same behavior is observed for the corresponding standard deviation 
(Fig.3c full circles) which is quite small and decreases when $T$ decreases. 
This shows firstly that with increasing temperature the SiO$_4$ 
tetrahedra survive even in the liquid phase but become more and more 
distorted and secondly that the glass transition does not strongly 
affect the local environment around the silicon atoms. On the
contrary the glass transition is more clearly visible in the angle Si-O-Si, 
which measures the relative position and orientation of two neighboring 
SiO$_4$ tetrahedra, as can be seen in Fig.3b. With decreasing temperature within the liquid phase,
the Si-O-Si angle increases and seems to converge towards $180^\circ$ (a least square quadratic
fit of the four points in the liquid phase gives an extrapolated value 
of 175$^\circ$), 
but again below $4000$ K this increase slows down 
and finally the angle converges towards a value close to $150^\circ$,
a value coherent with previous simulations (Vollmayr et al. 1996; 
Taraskin and Elliott 1997b), but 
slightly higher than the value $144^\circ$ found in X-ray diffraction 
experiments (Neuefeind and Liss 1996). The decrease of this angle 
is coherent with the 
views of a densifying network with increasing temperature this densification
taking place around the oxygen atoms. Also since it 
measures the relative orientation between two neighboring tetrahedra, this 
decrease corresponds to a decrease of the effective volume of the 
oxygen atoms with increasing temperature and since we work at constant 
volume it implies an expansion of the silicon volume, which is indeed
the behavior observed in Fig. 2. Concerning the standard deviation
of the Si-O-Si angle represented in Fig.3c (open circles) it increases 
rapidly in the liquid phase (above $4000$ K) while in the glass phase 
this increase is more slow. This again is an illustration of the 
structural freezing below $T_g$.\\
Another quantity which reflects the glass transition
is the effective coordination number, $z$, which is in fact the average
number of faces of the Vorono\"\i\ cells. This quantity should be considered
 with some care
since, generally, it does not correspond to the true ``chemical'' 
coordination number.
To give an example,  in $\beta$-cristobalite the
Vorono\"\i\ cell of the silicon atoms is a tetrahedron ($z=4$) while the
Vorono\"\i\ cell of the oxygen atoms is the polyhedron represented in
Fig. 4 with $z=8$. In the Vorono\"\i\ ``sense'' the nearest neighbors
of an oxygen atom in $\beta$-cristobalite are two silicon atoms (represented
by the triangular faces in Fig. 4), and six oxygen neighbors (represented
by the pentagonal faces in Fig. 4). 
Such a cell is  a truncated double-tetrahedron and is, in fact,
highly ``degenerate'':  some of its edges are shared by four neighboring
cells (instead of three usually). An infinitely small random perturbation
of the atomic positions ($\approx 10^{-2}$\AA ) is sufficient to create new 
extra-small triangular faces and the coordination number of the  
oxygen atoms flips to $z\approx 19.7$ 
while the coordination of the silicon atoms remains close to $4$.
This could be avoided by doing a smoothed 
Vorono\"\i\ tessellation which excludes small faces with an area smaller
than 10\% of the largest cell face (Rustad et al. 1991) 
but we have rather chosen not to use
this technique and analyze raw data. The variation of $z$ as a function
of temperature is represented in Fig. 5a for the silicon atoms and
Fig. 5b for the oxygen atoms. The glass transition can again clearly be
identified: for the silicon atoms, $z$ increases with increasing temperature
with a slope that is more important above $4000$ K. For the oxygen atoms, 
$z$ is approximately constant in the glass phase and then decreases for 
temperatures above $T_g$. It should be noted that the relative variation
of $z$ between $0$ and $7000$ K is small, especially for the oxygen atoms.
In the liquid phase, with decreasing temperature, the local structure around 
the silicon atoms evolves towards a  perfect tetrahedral arrangement 
($z=4$) and then converges to a value higher than $4$ in the glass phase due
again to the structural freezing below $T_g$.

\section{Discussion}

All these results can be discussed in the light of the conclusions drawn
for the analogous geometrical analysis performed in a monoatomic soft-sphere 
glass (Jund et al. 1997a, 1997b).
In that case, the system tries to reach an icosahedral arrangement 
(with dodecahedral Vorono\"\i\ cells) when
the temperature is lowered down from the liquid phase, but, since such an 
arrangement can not be realized at large distances in the regular 
three-dimensional space, the system gets frozen in a glass phase below a 
characteristic glass temperature.
Such geometrical frustration effects were in that case
the consequence of the degeneracy between the face-centered-cubic (FCC) and the hexagonal-close-packed (HCP) structures (Sadoc and Mosseri 1997).\\
Since all the standard deviations presented here seem to go to zero with 
decreasing temperature, when extrapolated
from the liquid phase, it is reasonable to assume that in the case of silica also a $T=0$ unreachable ideal local structure exists.
It is also reasonable to assume that such an ideal
arrangement corresponds to a perfect tetrahedral order for the four oxygens 
bounded to a given silicon atom, as it is for almost all of the known 
crystalline structures of silica. This assumption is
supported by the behavior of z, which tries to extrapolate to 4 when $T$ is 
lowered from the liquid phase, and it is not incompatible with our results 
for the variation of the angle O-Si-O with temperature (see Fig.3a). 
Even if this angle does not show a major change of behavior at $T_g$, 
an extrapolated value of 109.$^\circ$47 at $T=0$ is not inconsistent 
with the reported data above $T_g$. Moreover, since the angle Si-O-Si 
seems to extrapolate to 180$^\circ$, one can imagine that the ideal 
structure is made of tetrahedral units, like the
{\em sp}$_3$ coordination of carbon where the silicon atoms would be 
located at the carbon places and oxygen atoms
located in the middle of the C-C bonds (see Fig. 6a).
Indeed the tendency to build such a local structure should result from the 
form of the potential. In particular the tetrahedral arrangement of the 
oxygens around a silicon atom results from a combination of the 
Si-O attraction and the repulsion between the oxygens. The tendency to 
align the Si-O-Si bridges between neighboring tetrahedra is more 
subtle however, since the long range nature of the ionic part of the
potential certainly plays a role.\\
It is interesting to notice that, among all the known crystalline
structures of silica, two particular structures (at least) fully
satisfy these criteria, namely the $\beta$-cristobalite and
the tridymite structures. In these two structures  the above defined
tetrahedral units are stacked with sequences
ABCABC...(Fig.6b) and ABABAB...(Fig.6c), respectively, like in FCC and HCP 
structures. In fact these structures can be simply built from FCC and HCP 
structures, by adding to the original structure another one shifted by a 
fourth of the diagonal of the cubic cell (in the FCC case) and
by 3/8 of the $c$-axis of the hexagonal cell (in the HCP case). 
They correspond respectively to the diamond and wurtzite structures of carbon.
When these two structures are 
considered with the same density, they have exactly the same Si-O distance 
$d_{SiO}$, and therefore, in the two cases the Vorono\"\i\ cells for the 
silicon atoms are regular tetrahedra with the same volume 
$V_{Si}=\sqrt{3}d_{SiO}^3$. The  volume of the 
oxygen cells, $V_O$, is also the same and therefore the
two structures are characterized by the same Si/O volume ratio 
$R=V_{Si}/(2V_O)=9/55=0.164$, independent on
the density. In Fig.7 we have plotted $R=V_{Si}/(2V_O)$
as a function of $T$, as calculated from our simulations,
and reported the value $R=0.164$ at $T=0$ (open circle). When decreasing the 
temperature from
the liquid phase $R$ decreases as if it would like to reach a value quite 
close to (or even lower than) $0.164$. 
Therefore, one could conclude that may be the ideal structure
that the system would try to realize when lowering temperature
would be a mixture of tridymite and cristobalite phases. 
In fact this is not a satisfying solution since these two phases can 
very well arrange themselves under the form of crystallites separated by 
dislocations or stacking faults, and therefore crystallization should 
actually occur in our system. \\

One can find a solution by pursuing the analogy with soft sphere glasses 
or hard sphere packings for which also two degenerate ground state 
arrangements FCC and HCP exist.
In that case the system gets ``frustrated'' by trying to reach a 
local icosahedral structure (Jund et al. 1997a, 1997b; Jullien, Jund, 
Caprion and Quitmann 1996) which does not exist in 
the flat regular space.
Similarly to these model systems, the frustration can be resolved
by considering in our case a curved space with positive curvature, namely the 
sphere $S_3$ (Sadoc and Mosseri 1997). Consider a regular tetrahedron with a 
silicon atom in the middle and oxygen
 atoms at the centers of the faces. This tetrahedral unit contains one 
silica molecule.
One can exactly tile an $S_3$ space with 600 tetrahedra 
like this, and the resulting SiO$_2$ structure (which contains 600 
silicons and 1200 oxygens) satisfies
all the local requirements defined above. The 120 vertices of the unit 
tetrahedra are located on  the so-called
\{3,3,5\} polytope (Sadoc and Mosseri 1997). 
Note that an important difference with tridymite and cristobalite is that
this ideal structure is made of five-membered ..Si-O-... rings, instead 
of six-membered ones, since there are five unit tetrahedra around a 
tetrahedron edge in the \{3,3,5\} polytope.
It is not trivial to calculate the exact value of the ratio $R$ for this 
structure but a good approximation can be obtained by considering 
the tetrahedron unit in the flat three dimensional
space tangent to the curved space at the tetrahedron center.
 The silicon Vorono\"\i\ cell, limited by the bisector planes of the 
Si-O bonds, is a tetrahedron of volume 1/8 of the volume of the unit. 
Therefore the remaining volume for the two oxygen atoms is 7/8 and 
consequently $R=1/7=0.143$.
This value is only approximate due to the fact that volumes do not exactly
scale as distance-cubes in the curved space $S_3$ of positive radius $r_s$. 
An estimation of the error made can be obtained by performing the same 
reasoning for two concentric spheres of radii $r$ and $r/2$, for which one 
knows analytic formulae for their volumes (Sadoc and Mosseri 1997). 
Performing an expansion in $r/r_s$, one gets
$R\approx (1/7) (1+ (2/35)(r/r_s)^2 ...)$ instead of 1/7. Taking for $r$ 
the radius of the sphere of same volume than the one of the
unit tetrahedron (which is the volume of $S_3$ divided by 600), this formula
gives 0.144 instead of 0.143. Hence, in view of the roughness of the
extrapolations done, such an  error can reasonably be neglected.\\
This value is represented by the open square in Fig.7. It appears that this 
value corresponds to a better extrapolation of the four points above 
4000K than the one obtained from tridymite or cristobalite (the second order
fit of the four liquid points leads to a value of 0.136). 
Therefore one is tempted to conclude as in the soft sphere case 
(Jund et al. 1997a; 1997b): 
when lowering the temperature from the liquid phase the system evolves as if
it would try to build locally such an ideal structure, but since this structure cannot be realized in the regular
three dimensional space, the systems gets frozen in a glassy state below $T_g$.
Obviously, the above interpretation should be considered as a 
suggestion and there are certainly other ways to resolve the frustration.
One way is to consider the ``polytope 240'' described in chapter 2 
of the Sadoc and Mosseri book (Sadoc and Mosseri 1997). In that case, one 
would get six-membered rings. One could also imagine solutions with 
seven membered rings. The only argument in favor of our simple suggestion 
is the nice extrapolation of the volume ratio data to 0.143. We have 
checked that this ratio is significantly larger for the polytope 240 solution. Definitely more calculations are needed to validate our proposal. In 
particular a systematic enumeration of the $n$-membered rings as a 
function of temperature would be helpful.
Such an enumeration already exists in the literature (Vollmayr et al. 1996) 
but has been done as a function of quenching rate.

\section{ Conclusions }

With the use of classical molecular dynamics simulations combined with
the Vorono\"\i\ tessellation we have studied the evolution of the 
local structure around the particles of a model silica glass as a function
of temperature. The glass transition temperature, $T_g$, obtained from 
the variation of the potential energy versus the temperature is coherent with 
the value expected for systems quenched at $2.3\times10^{14}$ K/s. This
transition is clearly visible in the evolution of all the geometric 
characteristics of the Vorono\"\i\ cells. The study of the local density
fluctuations clearly demonstrates  the saturation of these fluctuations 
below $T_g$ which is the usual signature of the glass transition. The
study of the tetrahedral O-Si-O angle shows that with increasing temperature 
the SiO$_4$ tetrahedra survive but become more and more distorted. The 
variation of the angle Si-O-Si between two corner-sharing tetrahedra 
is coherent with the views of a densifying network with increasing temperature, this densification happening around the oxygen atoms, while a volume expansion
occurs around the silicon particles since we are considering microcanonical
ensembles. The different geometric characteristics and in particular
the average coordination number evolve with decreasing temperature as if
the system would like to reach an ideal 
structure at $T=0$, which cannot be realized due to geometrical
frustration. This evolution is frozen below the glass transition temperature,
and finally the system converges towards an amorphous structure.\\
Our results show that the use of the Vorono\"\i\ cell characteristics
gives not only useful informations on the local structure but can also 
be used to determine the glass transition unambiguously. This means that
even if nothing {\em dramatic} happens at the glass transition concerning
the local structure, {\em something} happens, which is basically a dynamic
freezing of the natural evolution of the structure towards an unreachable ideal structure.
Of course we are far away from the timescales used in experiment,
but this study, together with others, permits to investigate what happens
on the microscopic level in an attempt to explain what is observed
at the macroscopic level.


\section*{ Acknowledgments }
We thank C. Levelut and J. Pelous for helpful discussions.
Part of the numerical calculations were done at CNUSC
(Centre National Universitaire Sud de Calcul), Montpellier.


\newpage
\begin{center}
REFERENCES
\end{center}
\vskip 1cm

\hspace*{-0.63cm}ACHIBAT, T., BOUKENTER, A. and DUVAL, E., 1993, 
{\em J. Chem. Phys.}, 
{\bf 99}, 2046.\\
AKKERMANS, E. and MAYNARD, R., 1985, {\em Phys. Rev. B}, {\bf 32}, 785.\\
ALLEN, M.P. and TILDESLEY, D.J., 1990, 
{\em Computer simulation of liquids}, Oxford University Press, New-York.\\
ANGELL, C.A., 1988, 
{\em J. Chem. Phys. Solids}, {\bf 49}, 863.\\
BERMEJO, F.J., CRIADO, A. and MARTINEZ, J.L., 1994, 
{\em Phys.\ Lett. A}, {\bf 195}, 236.\\
B\"ORJESSON, F.L., HASSAN, A.K., SWENSON, J., TORELL, L.M. and FONTANA, A., 1993, {\em Phys.\ Rev.\ Lett.}, {\bf 70}, 1275.\\
BUCHENAU, U., ZHOU, H.M., N\"{U}CKER, N., GILROY, K.S. and PHILLIPS, W.A.,
1988, {\em Phys.\ Rev.\ Lett.}, {\bf 60}, 1318.\\
DELLA VALLE, R.G. and ANDERSEN, H.C., 1992, 
{\em J. Chem. Phys.}, {\bf 97}, 2682.\\
ELLIOTT, S.R., 1991, {\em Phys.\ Rev.\ Lett.}, {\bf 67}, 711.\\
EWALD, P.P., 1921, 
{\em Ann. Phys.}, {\bf 64}, 253.\\
FAYOS, R., BERMEJO, F.J., DAWIDOWSKI, J., FISCHER, H.E. AND GONZALEZ, M.A., 
1996, {\em Phys.\ Rev.\ Lett.}, {\bf 77}, 3823.\\
GASKELL, P.H. AND WALLIS, D.J., 1996, {\em Phys.\ Rev.\ Lett.}, {\bf 76}, 66.\\
GELLATLY, B.J. and FINNEY, J.L., 1982, 
{\em J. Non Cryst. Solids}, {\bf 50}, 313.\\
GUILLOT, B. and GUISSANI, Y., 1997, 
{\em Phys.\ Rev.\ Lett.}, {\bf 78}, 2401.\\
GUREVICH, V.L., PARSHIN, D.A., PELOUS, J., and SCHOBER, H.R., 1993, 
{\em Phys. Rev. B}, {\bf 48}, 16318.\\
HOLENDER, J.M. and MORGAN, G.J., 1991,
{\em J. Phys: CM}, {\bf 3}, 1947.\\
HORBACH, J., KOB, W. and BINDER, K., 1997, preprint.\\
JIN, W., VASHISTA, P., KALIA, R.K. and RINO, J.P., 1993, 
{\em Phys. Rev. B} {\bf 48}, 9359.\\
JULLIEN, R., JUND, P., CAPRION, D. and QUITMANN, D., 1996, 
{\em Phys.\ Rev.\ E}, {\bf 54}, 6035.\\ 
JUND, P., CAPRION, D. and JULLIEN, R., 1997a, 
{\em Europhys. Lett.}, {\bf 37}, 547.\\
JUND, P., CAPRION, D. and JULLIEN, R., 1997b,
{\em Mol. Sim.}, {\bf 20}, 3.\\
JUND, P., CAPRION, D. and JULLIEN, R., 1997c,
{\em Phys.\ Rev.\ Lett.}, {\bf 79}, 91.\\
KLEMAN, M., 1983,
{\em J. de Physique (France)}, {\bf 44}, L295.\\
KLEMAN, M., 1985,
{\em J. de Physique (France)}, {\bf 46}, L723.\\
KLEMAN, M., 1989,
{\em Adv. in Phys.}, {\bf 38}, 605.\\
KLEMAN, M. and SADOC, J.-F., 1979,
{\em J. de Physique (France)}, {\bf 40}, L569.\\
MALINOVSKY, V.K., NOVIKOV, V.N., PARSHIN, P.P., SOKOLOV, A.P. and ZEMLYANOV, M.G., 1990, {\em Europhys. Lett.}, {\bf 11}, 43.\\
MARTIN, A. and BRENIG, W., 1974, {\em Phys. Status Solidi B}, {\bf 64}, 163.\\
NAKANO, A., BI, L., KALIA, R.K. and VASHISTA, P., 1993, 
{\em Phys.\ Rev.\ Lett.}, {\bf 71}, 85.\\
NEUEFEIND, J. and LISS, K.D., 1996, 
{\em Ber. Buns. Ges. Phys. Chem.}, {\bf 100}, 1341.\\
NOVIKOV, V.N. and SOKOLOV, A.P., 1991, {\em Solid State
Commun.}, {\bf 77}, 243.\\
RUSTAD, R., YUEN, D.A. and SPERA, F.J., 1991, 
{\em Phys. Rev. B} {\bf 44}, 2108.\\
SADOC, J.F. and MOSSERI, R., 1997, 
{\em La frustration g\'eom\'etrique}, Alea Collection, Eyrolles ed., Saclay, 
France.\\
SARNTHEIN, J., PASQUARELLO, A. and CAR, R., 1995, 
{\em Phys.\ Rev.\ Lett.}, {\bf 74}, 4682.\\
SERVALLI, G. and COLOMBO, L., 1993, 
{\em Europhys. Lett.}, {\bf 22}, 107.\\
SCHIRMACHER, W., and WAGENER, M., 1993, {\em Solid State Commun.}, 
{\bf 86}, 597.\\
SHUKER, E. and GAMMON, R.W., 1970, {\em Phys.\ Rev.\ Lett.}, {\bf 25}, 223.\\
SOKOLOV, A.P., KISLIUK, A., SOLTWISCH, M. and QUITMANN, D., 1992,
{\em Phys.\ Rev.\ Lett.}, {\bf 69}, 1540.\\
STILLINGER, F.H. and WEBER, T.A., 1982, 
{\em  Phys. Rev. A}, {\bf 25}, 978.\\
TARASKIN, S.N. and ELLIOTT, S.R., 1997a, 
{\em Phys. Rev. B} {\bf 55}, 1.\\
TARASKIN, S.N. and ELLIOTT, S.R., 1997b,
{\em Europhys. Lett}, {\bf 39}, 37.\\
TERKI, F., LEVELUT, C., BOISSIER, M. and PELOUS, J., 1996, 
{\em Phys. Rev. B} {\bf 53}, 2411.\\
VAN BEEST, B.W.H., KRAMER, G.J. and VAN SANTEN, R.A., 1990, 
{\em Phys.\ Rev.\ Lett.}, {\bf 64}, 1955.\\
VOLLMAYR, K., KOB, W. and BINDER, K., 1996, 
{\em Phys. Rev. B} {\bf 54}, 15808.\\
WOODCOK, L.V., ANGELL, C.A. and CHEESEMAN, P., 1976, 
{\em J. Chem. Phys.}, {\bf 65}, 1565.\\

%
\newpage

\begin{figure}
\vskip 1cm

Fig. 1. Average potential energy per particle as a function of temperature.
\\

Fig. 2. Variation of the cell volume versus temperature: (a) silicon;
 (b) oxygen.\\
(c) Standard deviation $\sigma_V$ versus temperature: $\bullet$: silicon; 
$\circ$: oxygen.
\\

Fig. 3. Variation as a function of temperature of: (a) $\theta$, the O-Si-O angle, (b) $\phi$, the Si-O-Si 
angle; (c) the standard deviations $\sigma_\theta$ and $\sigma_\phi$ 
versus temperature: $\bullet$: silicon; $\circ$: oxygen.\\

Fig. 4. The Vorono\"\i\ cell of an oxygen atom in $\beta$-cristobalite.
\\

Fig. 5. Variation of the effective coordination, $z$, as a function of
temperature: (a) silicon atoms; (b) oxygen atoms.
\\

Fig. 6. The ideal tetrahedral unit around a silicon atom (a) is shown 
together with the trydimite (b) and $\beta$-cristobalite (c) structures viewed 
from the top. In (b) and (c) two successive layers connected to a single top 
unit are represented (the positions of the oxygen atoms have been omitted).
\\

Fig. 7. Variation of $R = V_{Si}/2V_{O}$ as a function of temperature. 
$\circ$: value of $R$ obtained in tridymite or $\beta$-cristobalite; 
$\Box$: value of $R$ obtained in the ideal structure built on $S_3$. 
 
\end{figure}


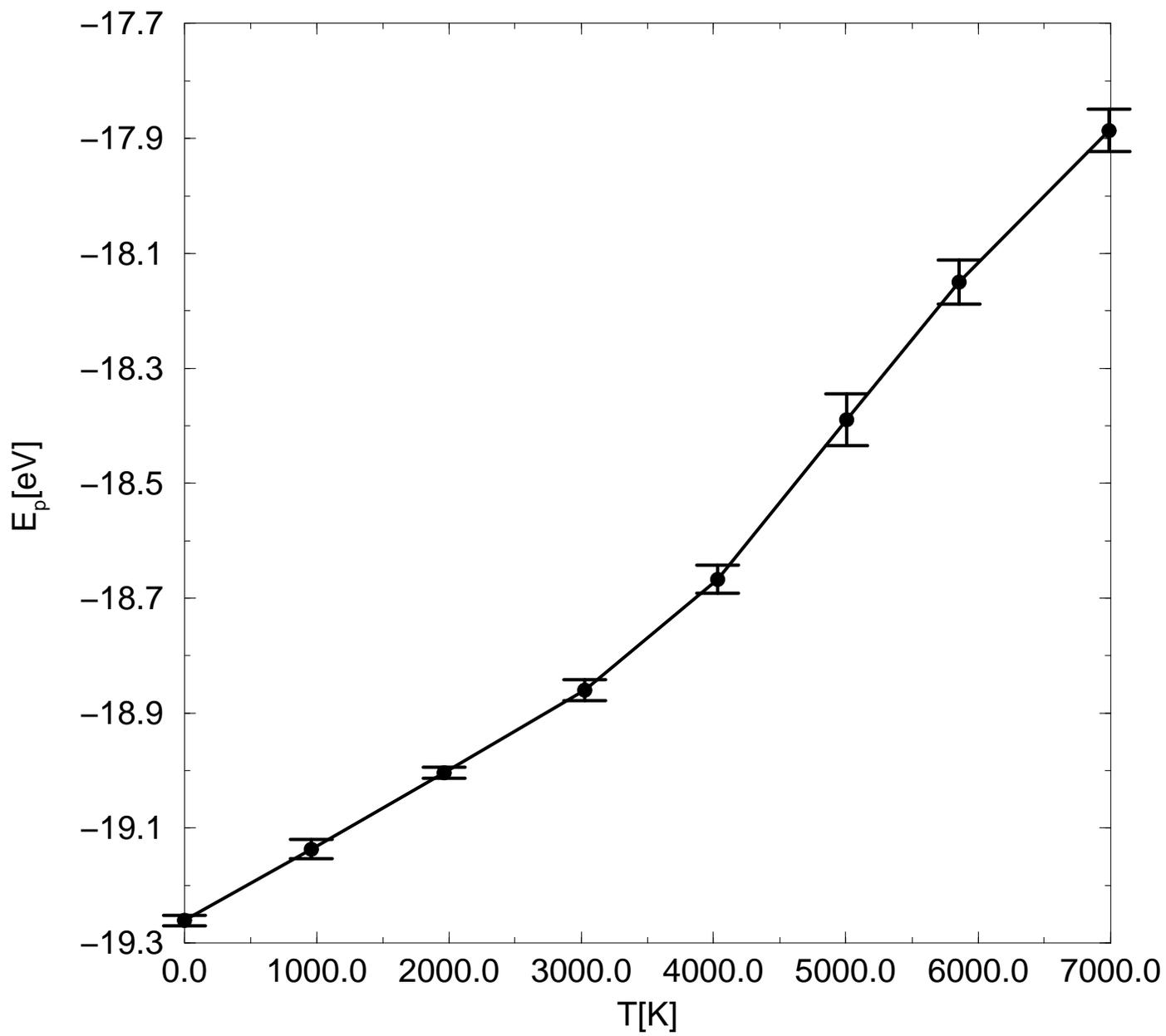
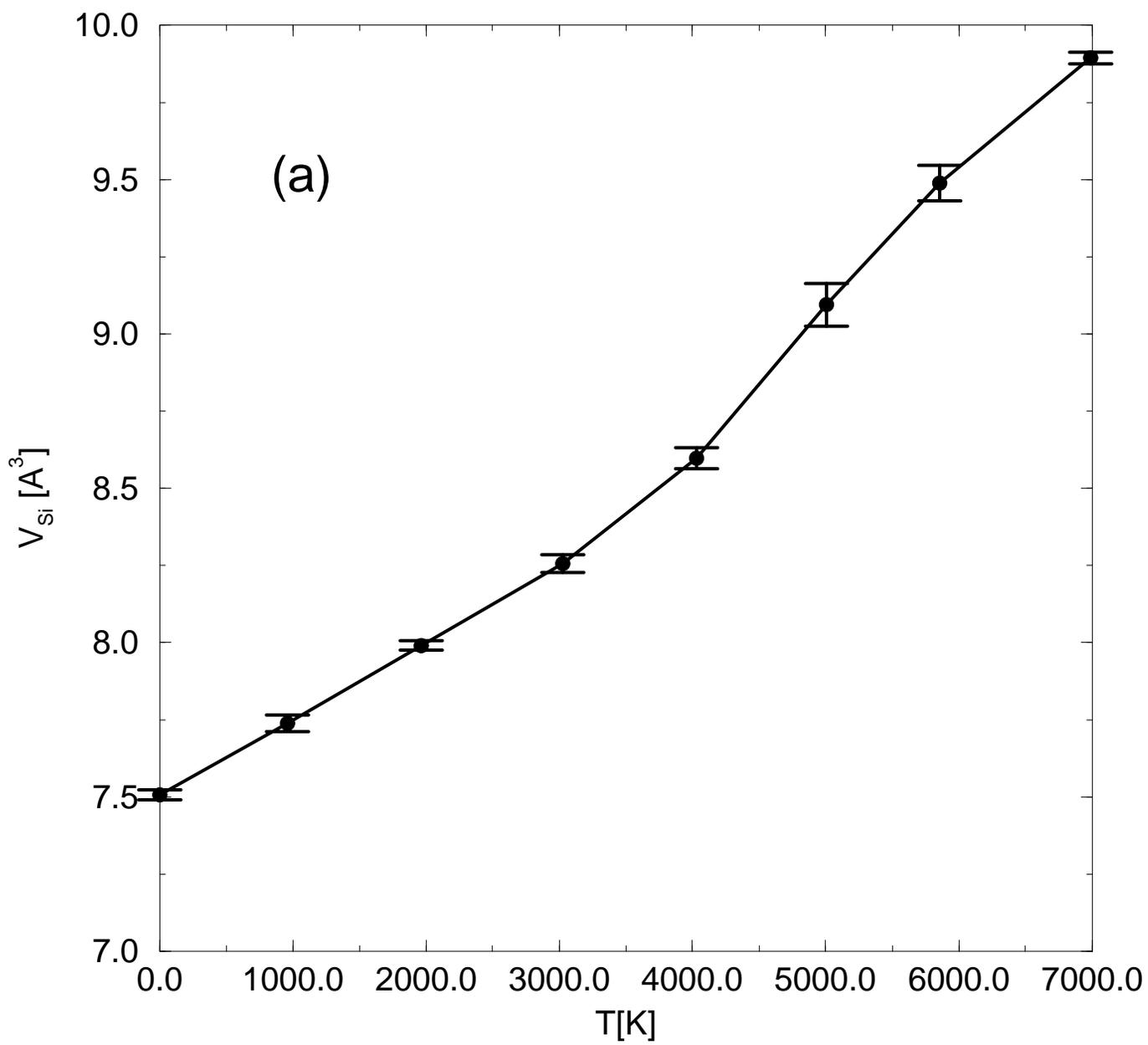
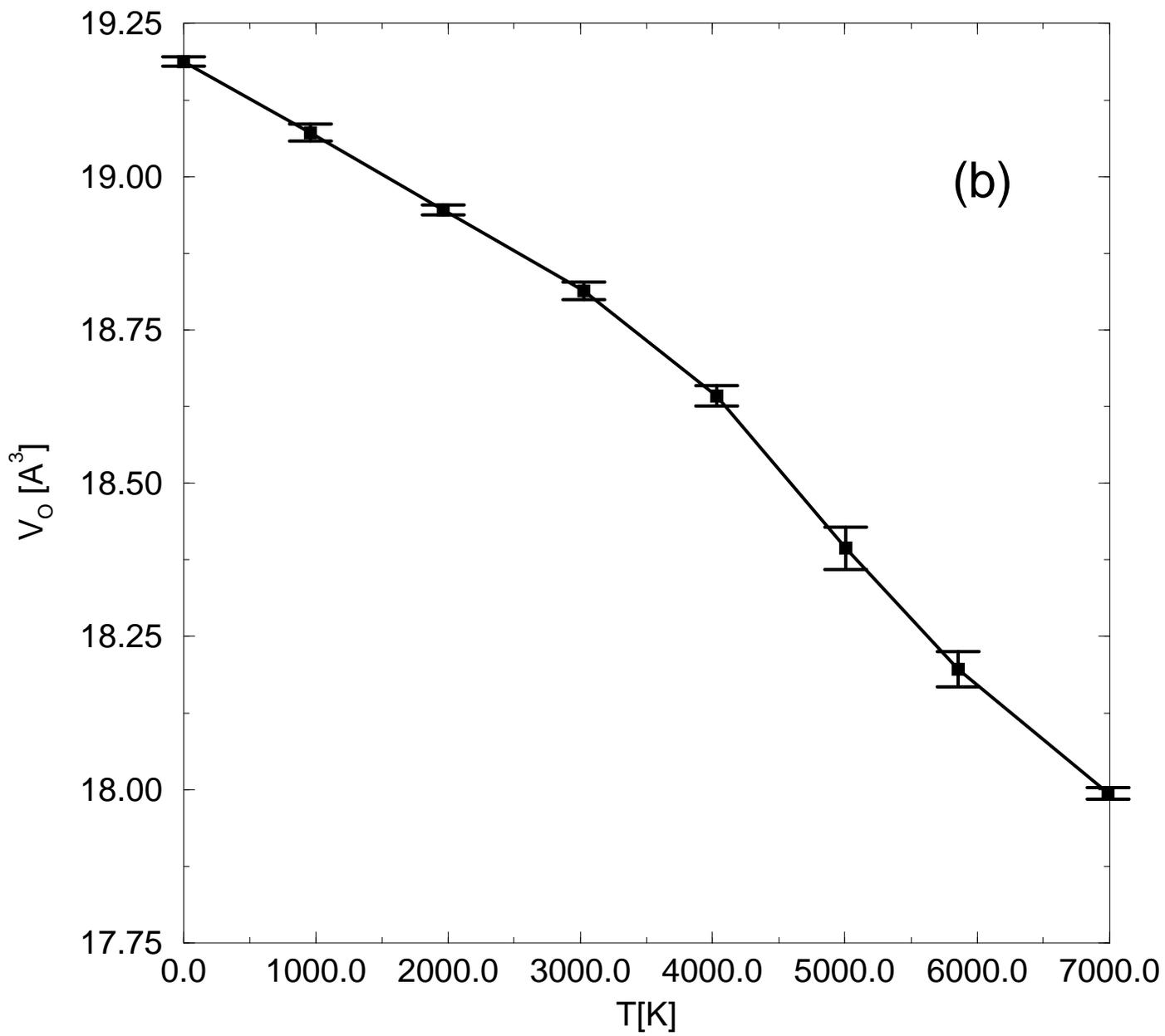
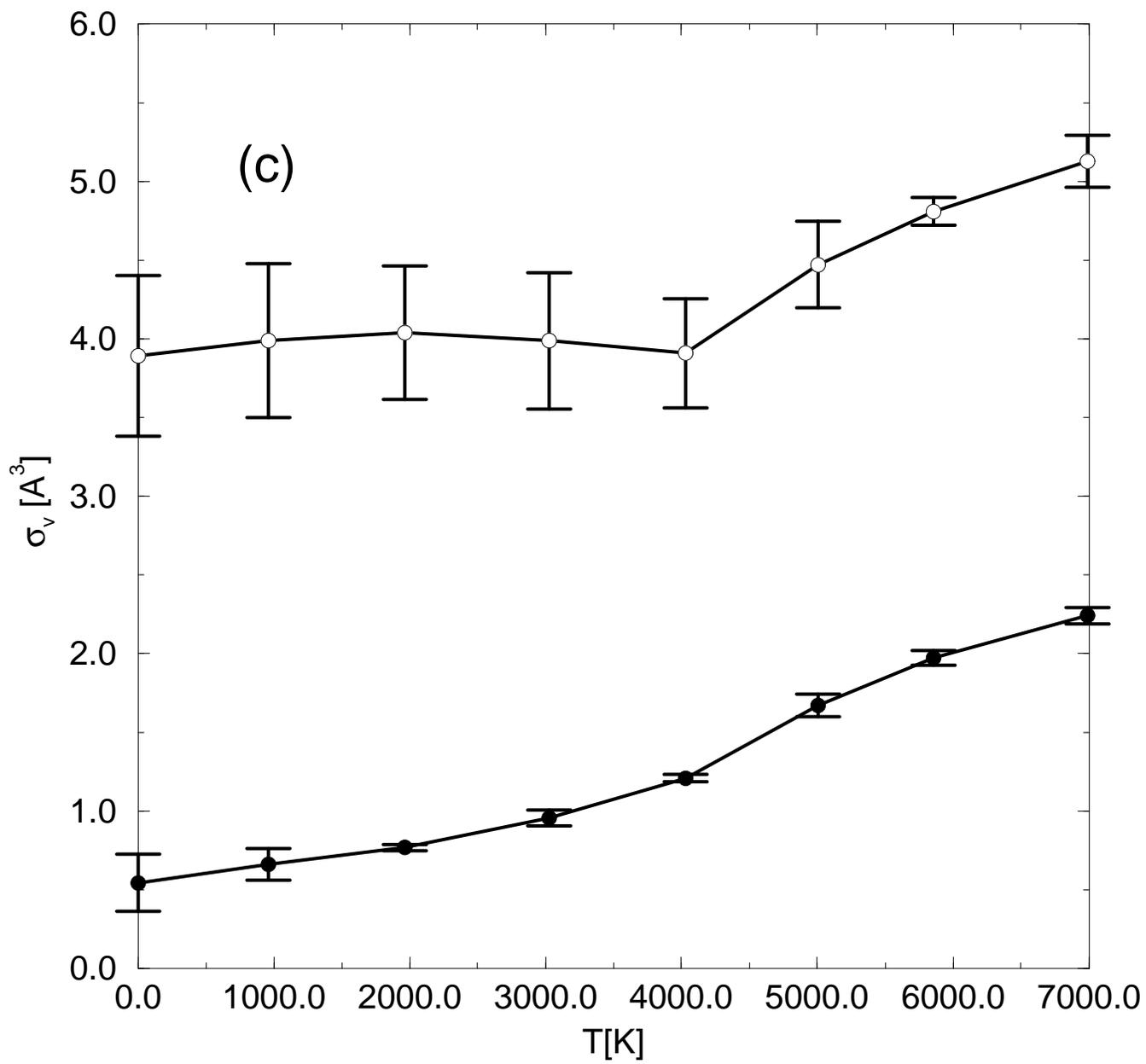
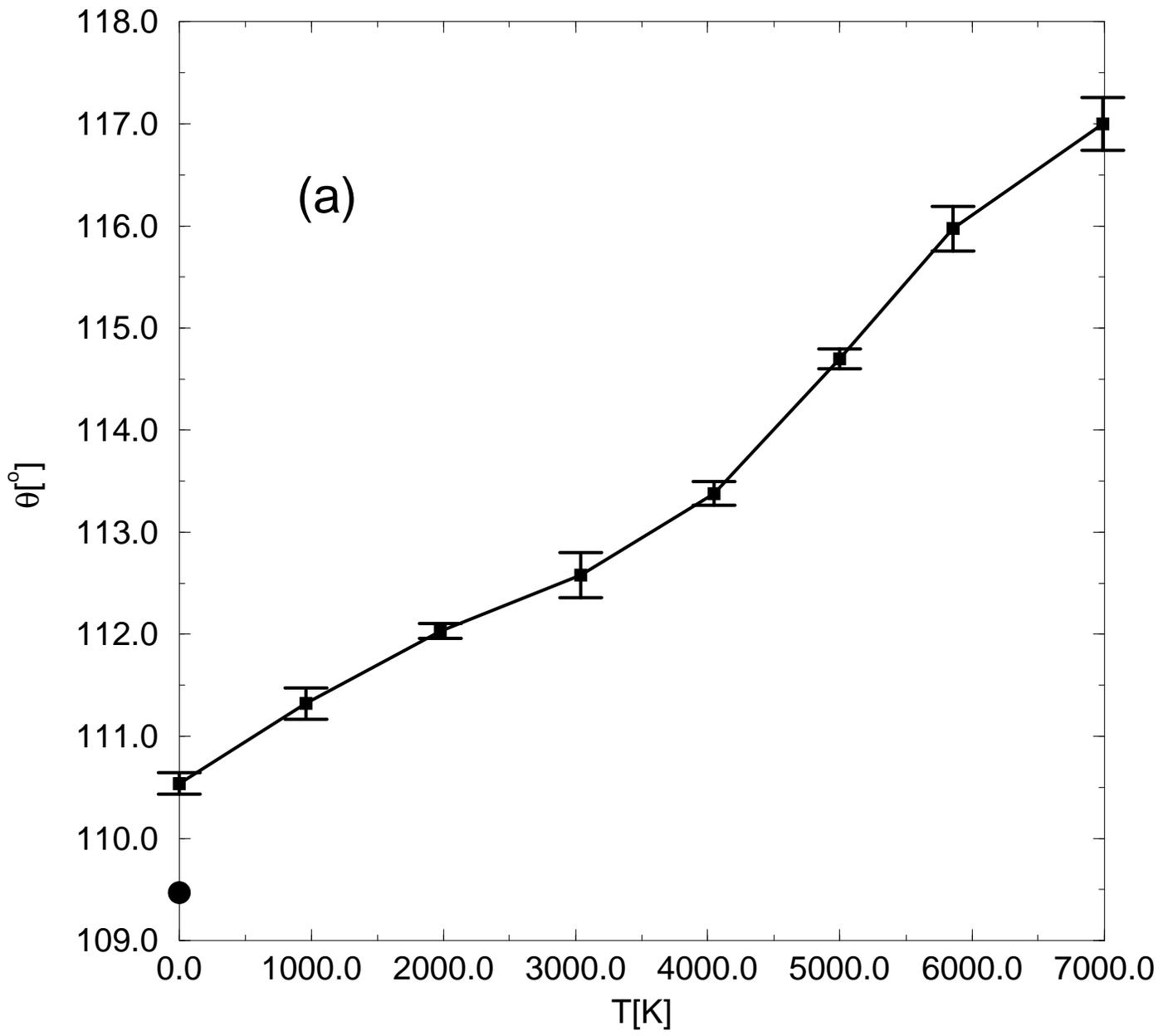
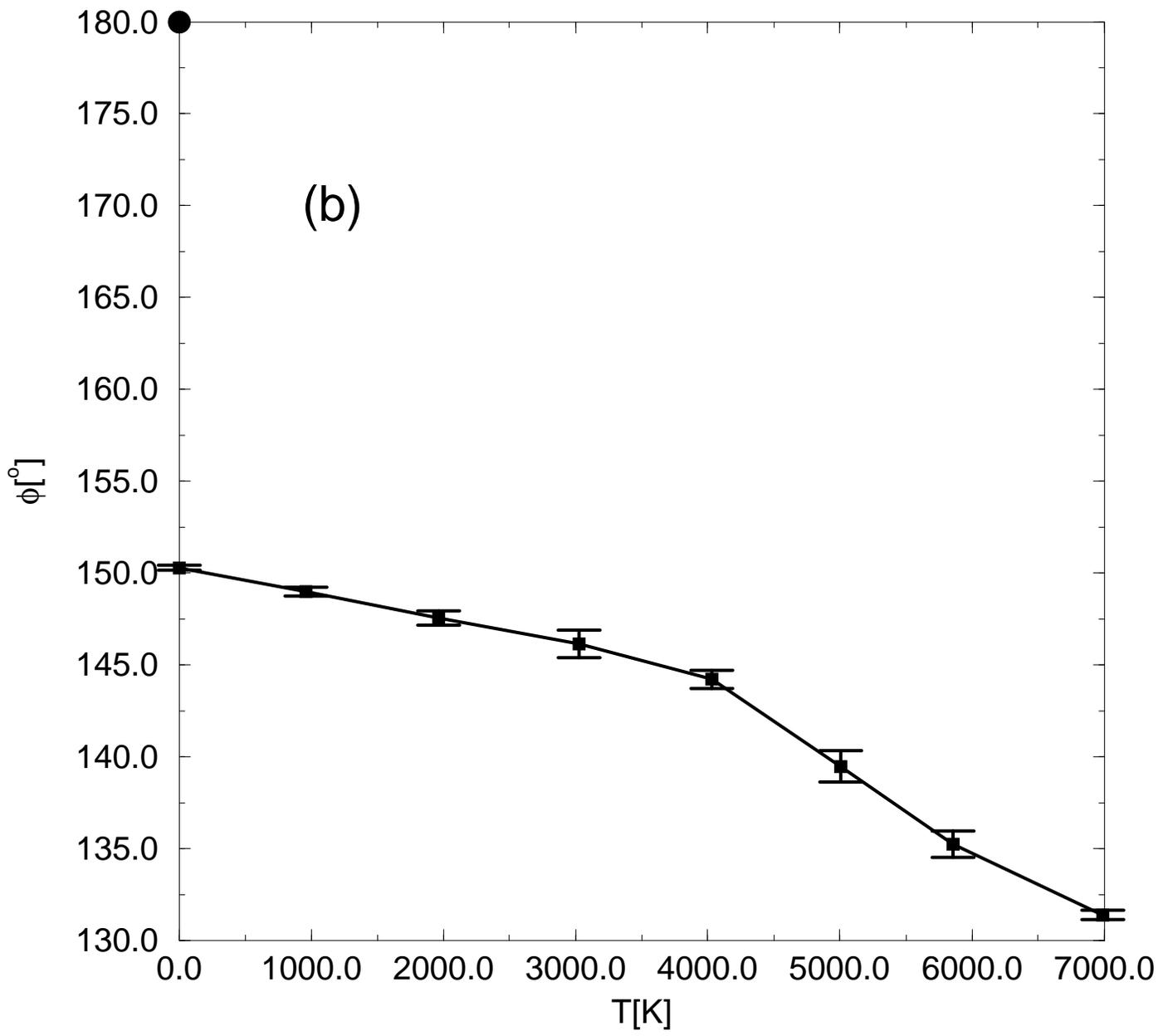
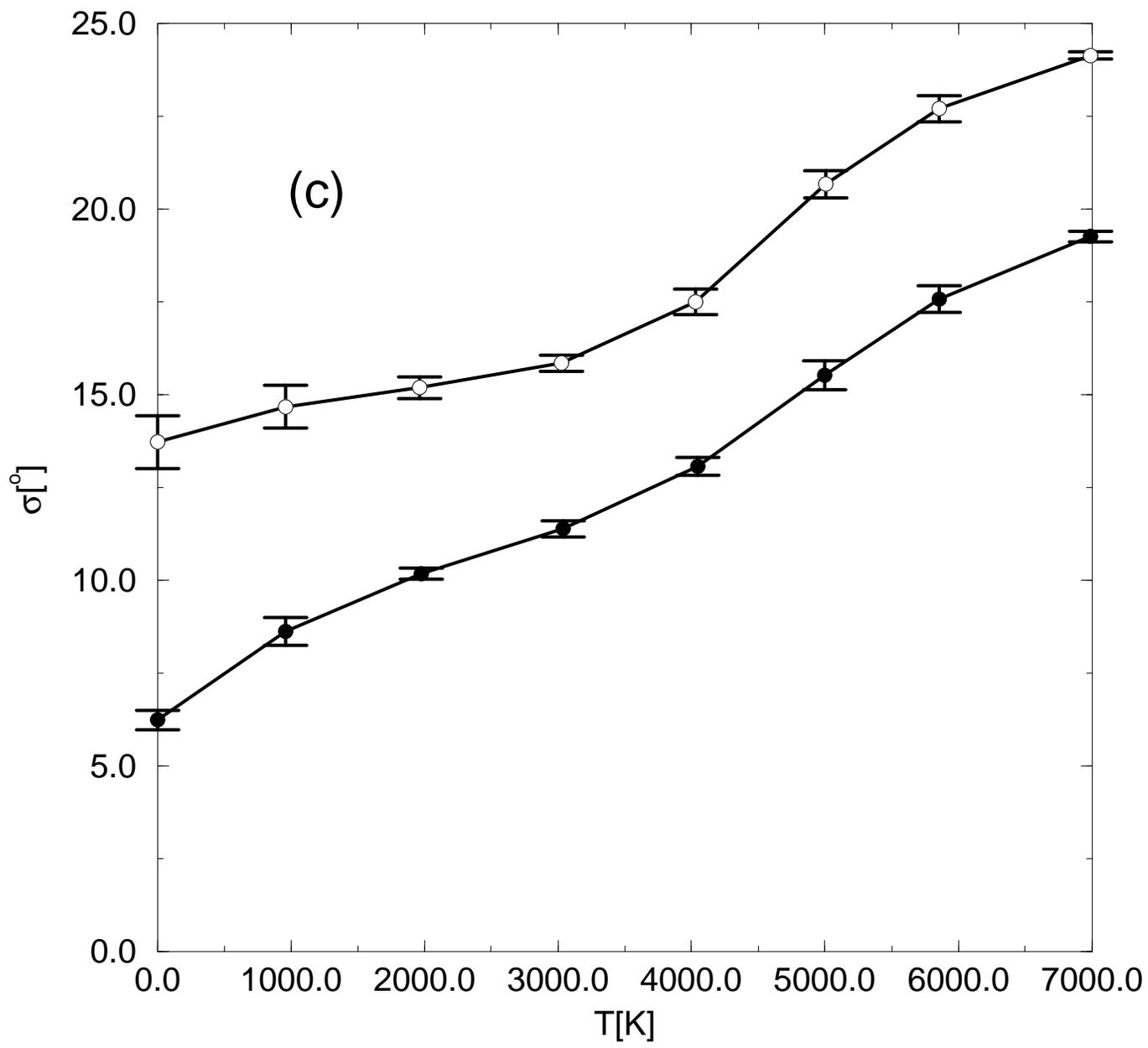
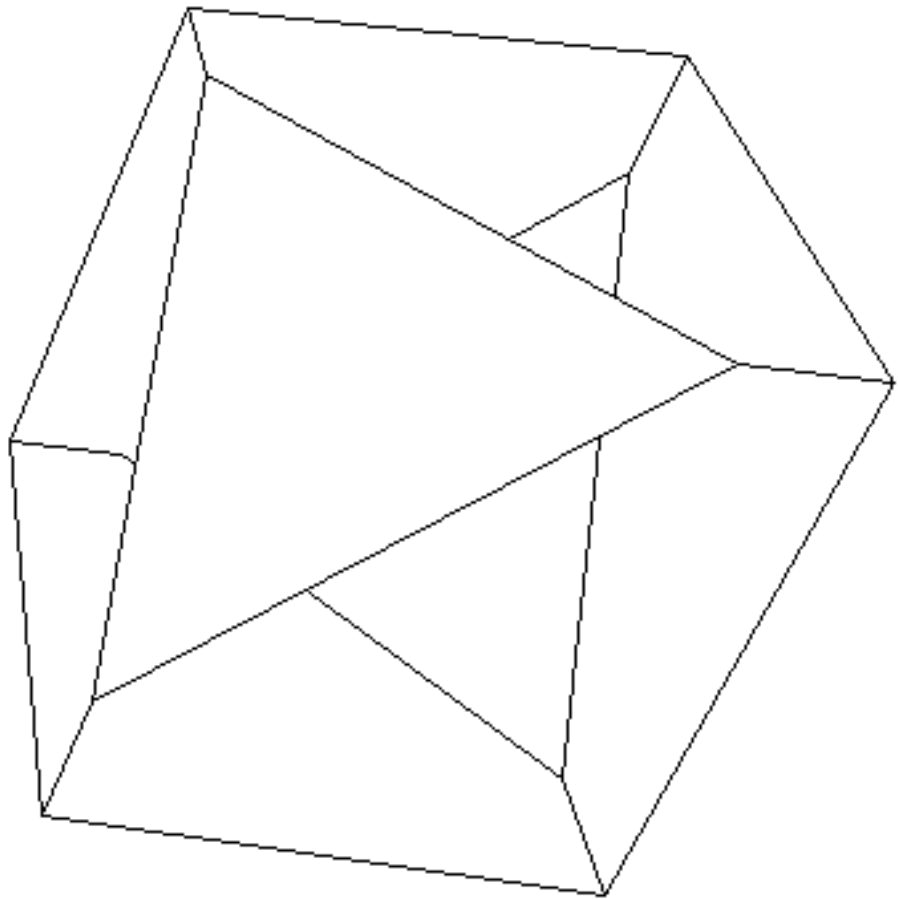
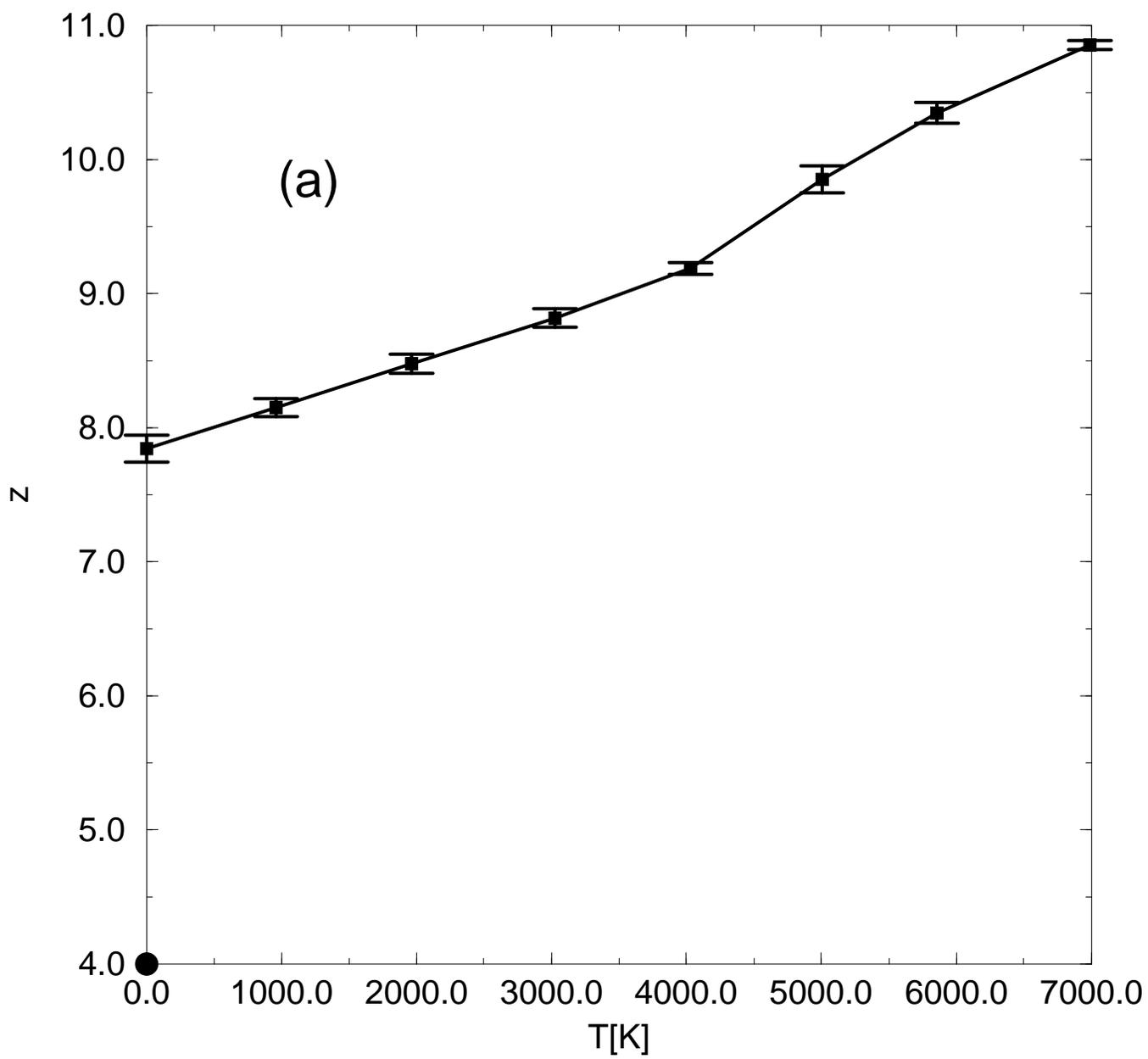
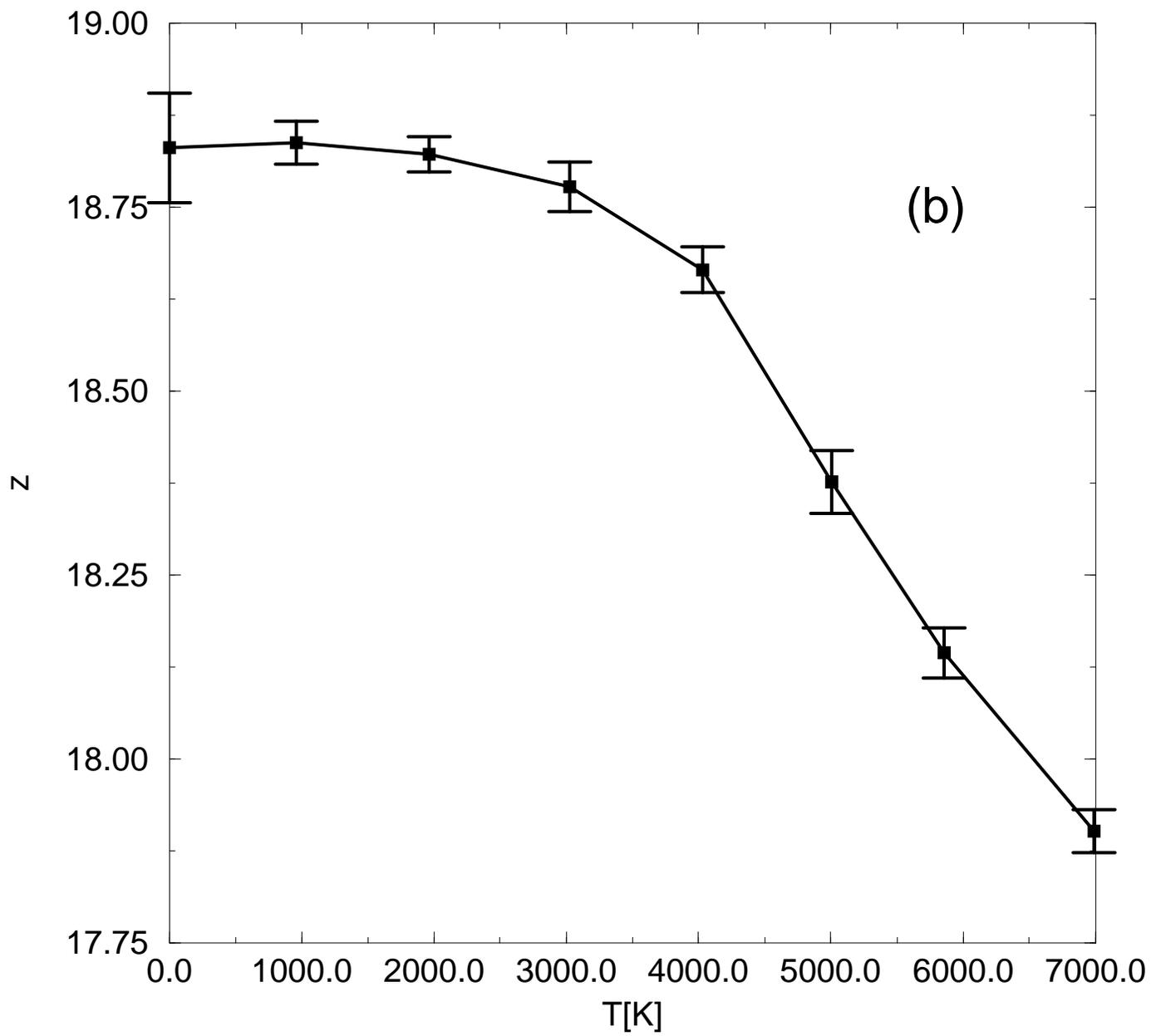
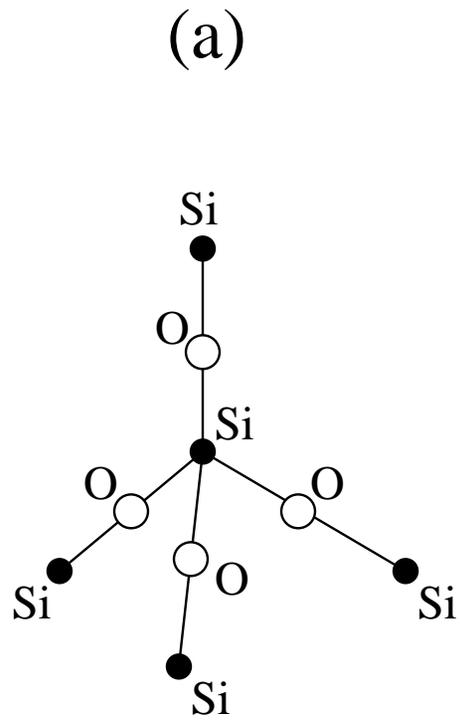
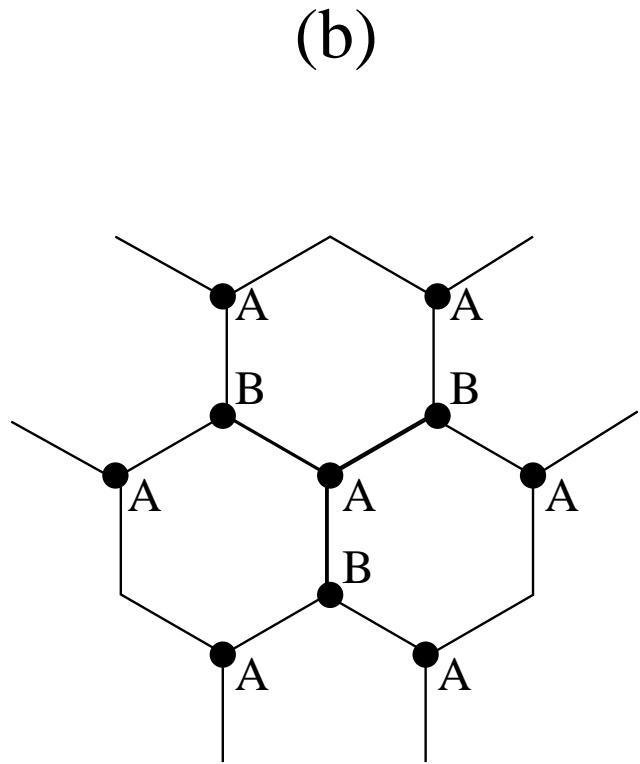
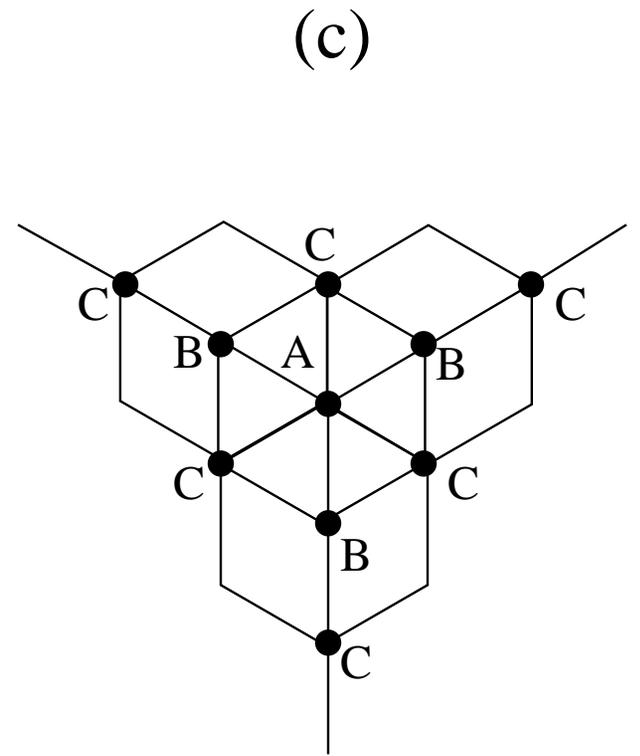
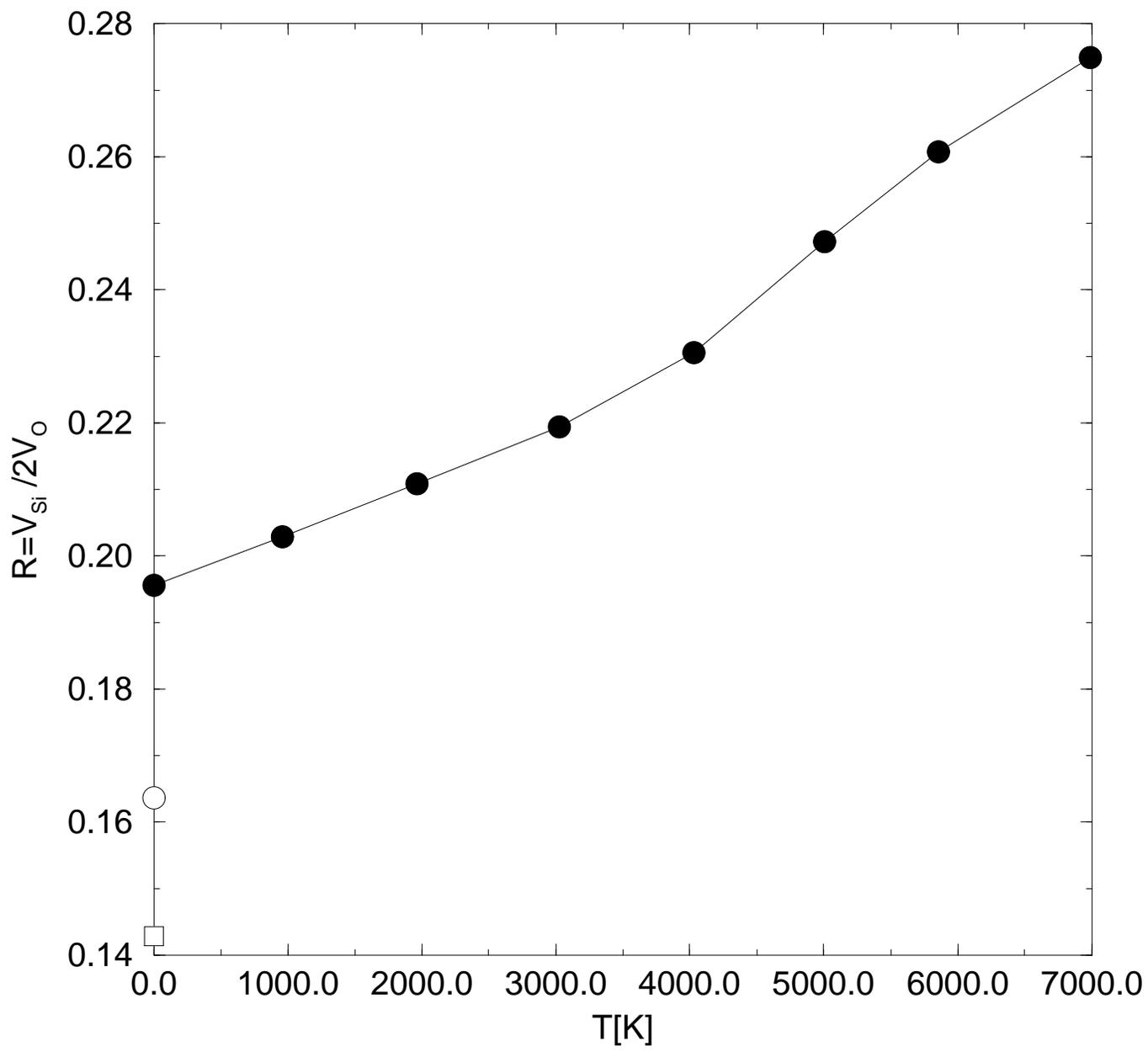
\end{document}